\documentclass[aip,jap,reprint,superscriptaddress,showpacs,floatfix]{revtex4-1}
\usepackage{amsmath}%
\usepackage{amsfonts}%
\usepackage{amssymb}%
\usepackage{graphicx}
\usepackage{color}
\usepackage{tabularx}
\usepackage{braket}

\begin{document}

\title{Hybrid quantum circuit with implanted erbium ions}

\author{S.~Probst}
\affiliation{Physikalisches Institut, Karlsruhe Institute of Technology, D-76128 Karlsruhe, Germany}
\author{N.~Kukharchyk}
\affiliation{Angewandte Festkörperphysik, Ruhr-Universit\"at Bochum, Universit\"atsstra\ss e 150, D-44780 Bochum, Germany}
\author{H.~Rotzinger}
\affiliation{Physikalisches Institut, Karlsruhe Institute of Technology, D-76128 Karlsruhe, Germany}
\author{A.~Tkal\v{c}ec}
\affiliation{Physikalisches Institut, Karlsruhe Institute of Technology, D-76128 Karlsruhe, Germany}
\author{S.~W\"unsch}
\affiliation{Institut f\"{u}r Mikro- und Nanoelektronische Systeme, Karlsruhe Institute of Technology, D-76189 Karlsruhe, Germany}
\author{A.~D.~Wieck}
\affiliation{Angewandte Festkörperphysik, Ruhr-Universit\"at Bochum, Universit\"atsstra\ss e 150, D-44780 Bochum, Germany}
\author{M.~Siegel}
\affiliation{Institut f\"{u}r Mikro- und Nanoelektronische Systeme, Karlsruhe Institute of Technology, D-76189 Karlsruhe, Germany}
\author{A.~V.~Ustinov}
\affiliation{Physikalisches Institut, Karlsruhe Institute of Technology, D-76128 Karlsruhe, Germany}
\affiliation{Laboratory of Superconducting Metamaterials, National University of Science and Technology ``MISIS'', Moscow 119049, Russia}
\author{P.~A.~Bushev}
\affiliation{Experimentalphysik, Universit\"at des Saarlandes, D-66123 Saarbr\"{u}cken, Germany}

\date{\today}

\begin{abstract}
We report on hybrid circuit QED experiments with focused ion beam implanted Er$^{3+}$ ions in Y$_2$SiO$_5$ coupled to an array of superconducting lumped element microwave resonators. The Y$_2$SiO$_5$ crystal is divided into several areas with distinct erbium doping concentrations, each coupled to a separate resonator. The coupling strength is varied from 5\,MHz to 18.7\,MHz, while the linewidth ranges between 50\,MHz and 130\,MHz. We confirm the paramagnetic properties of the implanted spin ensemble by evaluating the temperature dependence of the coupling. The efficiency of the implantation process is analyzed and the results are compared to a bulk doped Er:Y$_2$SiO$_5$ sample. We demonstrate the successful integration of these engineered erbium spin ensembles with superconducting circuits.
\end{abstract}


\maketitle


A future quantum information technology will most probably rely on employing different quantum systems, where each subsystem is specialized on fulfilling distinct tasks~\cite{Zoller2009, Nori2013}. For instance, modern superconducting (SC) quantum circuits are well suited for implementing scalable and fast quantum processors~\cite{Clarke2008}. However, these SC qubits suffer from relatively short coherence times~\cite{Martinis2010}. In contrast, spin doped solids possess long coherence times of up to a second~\cite{Tyryshkin2011} such that they can serve as a quantum memory. Hybrid circuit QED offers a promising way for implementing a complete quantum computer, i.e. a processor interfaced with a memory unit~\cite{Bertet2011}.

In the recent years, Y$_2$SiO$_5$ (YSO) crystals doped with rare-earth (RE) ions have moved into the focus of quantum information science~\cite{Gisin2010, Sellars2010, Gisin2011}. Additionally, strong coherent coupling of Er:YSO to a SC lumped element microwave resonator has been demonstrated~\cite{Probst_PRL13}. Most of the current research activity relies on RE:YSO crystals which are typically grown using the Czochralski method, where the RE doping takes place during the growth process~\cite{Usmani2012, Kroell2013, Riedmatten2014}.

In this article, we focus on the practical implementation of a hybrid quantum system. In a practical circuit, memory elements need to be placed at specific positions, where they can fulfill their tasks without interfering with the rest of the quantum circuitry. One possible approach is to locally implant spins into an empty crystal or directly into the substrate where the circuit is fabricated on \cite{kolesov2012}. Recently, weak coupling ($\sim$ 1~MHz) of a superconducting resonator to Gd$^{3+}$ ions implanted into a sapphire substrate has been reported \cite{Wisby2014}. Here, we employ YSO as a host material for RE ions, which promises long optical and spin coherence times \cite{Sun2006,Bertaina2007,Bertaina2009PRB}. Moreover, the optical transition of Er$^{3+}$ lies within the standard telecom C-band, which allows for the implementation of a reversible coherent microwave to optical interface for quantum communication \cite{Rabl2010,Tian2004}.

In our work, we use a focused ion beam (FIB) to implant Er$^{3+}$ ions into an undoped Y$_2$SiO$_5$ (YSO) crystal with high spatial resolution\cite{Kukharchyk2014}. We then perform circuit QED experiments on these crystals and confirm the successful implantation of erbium ions by studying the electron spin resonance properties at the single photon level.

\begin{figure}[ht!]
\includegraphics[width=1\columnwidth]{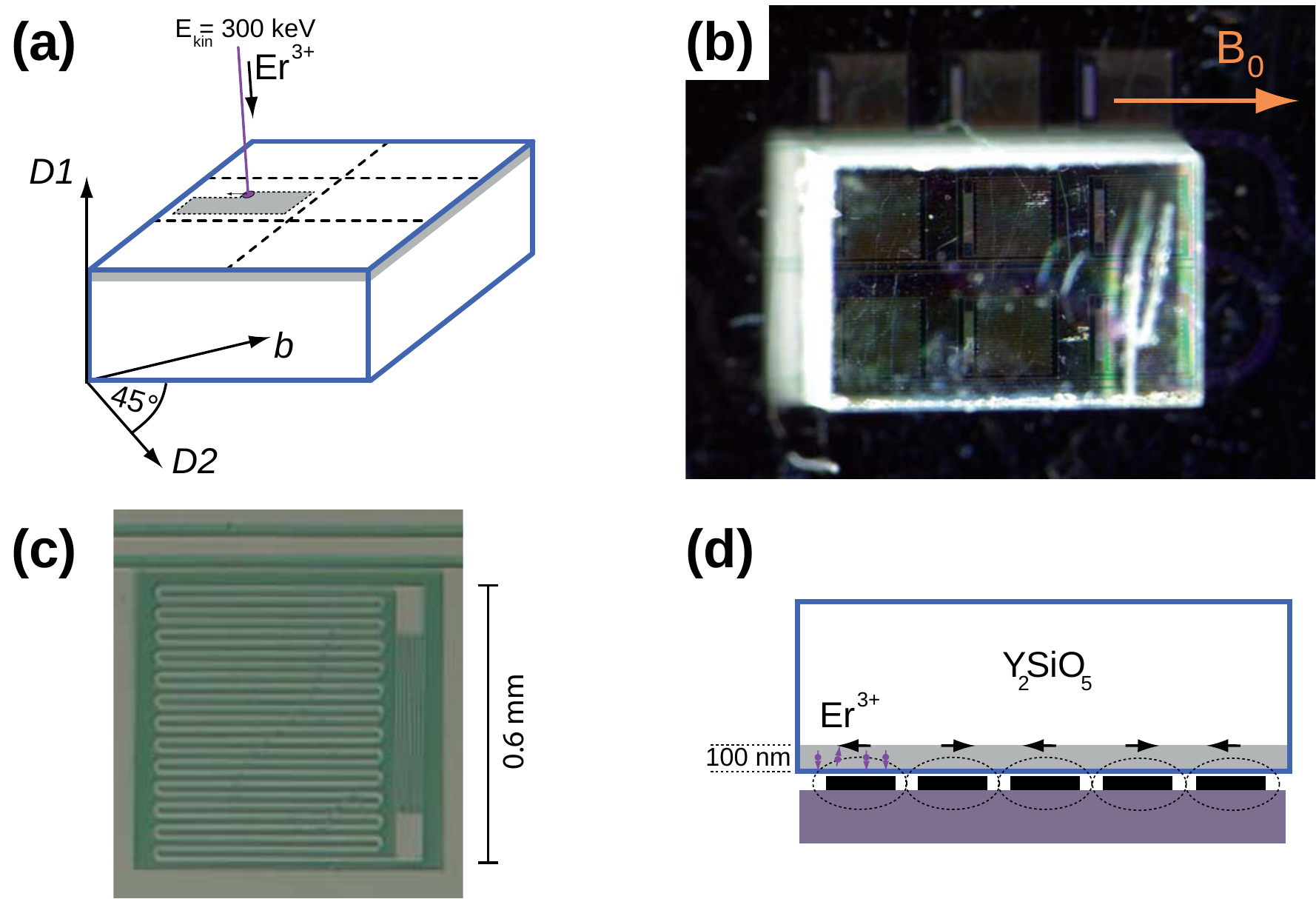}
\caption{(Color online) \textbf{(a)} Schematics of the focused ion beam implantation process and orientation of the YSO crystal. \textbf{(b)} Picture of the sample showing the doped crystal on top of the superconducting resonator array. \textbf{(c)} Optical micrograph of the superconducting LE resonator. \textbf{(d)} Schematics of the coupling geometry.}
\label{fig:Setup}
\end{figure}

Figure~\ref{fig:Setup}(a) shows a sketch of an implantation process. The crystal's surface is divided into several areas where we implant Er$^{3+}$ ions in order to study the effect of the amount of implanted ions per area (ion fluence) on the coupling strength and inhomogeneous spin linewidth. Erbium ions were implanted into the YSO crystals in an EIKO-100 FIB system with an energy of 300 keV. The ions were extracted from an Au$_{78.4}$ Er$_{10}$ Si$_{11.6}$ liquid metal ion source (LMIS), developed by A. Melnikov at al.\cite{melnikov2002}. The ion beam was accelerated in a potential of 100~kV and separated into its ion species by a built-in Wien filter.
The resolution of the filter does not allow for a fine separation between the Erbium isotopes, but it was possible to minimize the amount of the unwanted $^{167}$Er isotope to less than 5\%.
We estimate the amount of the ion fluence $Fl = {I  \frac{t}{S}} $, where $I$ is ion current, $t$ is the dwell-time of the beam on a single drawing point, and $S$ is the area covered by the ion beam.

As the ions penetrate the YSO crystal, they trigger a number of processes in the crystal, mainly ion-ion collisions resulting in lattice defects. The implantation process was simulated with the SRIM software~\cite{Ziegler2010}, which yields an ion distribution with a mean depth of 60~nm and 39~nm mean deviation~\cite{Kukharchyk2014}. Since the incident Er$^{3+}$ beam severely damages the crystal lattice, thermal annealing is employed for its restoration. We have performed both rapid thermal annealing (RTA) and $1.5-2$~hours long-time annealing.

There are two ways to investigate the properties of implanted ions, either optically, i.e. by confocal photoluminscence~\cite{Kukharchyk2014}, or in the microwave frequency range using electron spin resonance (ESR). In the case of a magnetic ion, ESR allows to obtain information about the properties of spins in solids, such as g-factors, relaxation and coherence times, and inhomogeneous linewidth~\cite{SchweigerESR}. Typically, 3D resonators deployed in ESR measurements, probe the spins distributed over the whole crystal volume. In contrast, on-chip ESR allows to study the surface region of the paramagnetic sample~\cite{Shuster2010, Bushev2011, Bertet2012, Morton2013}. Since the implanted ions only occupy a region 100~nm below the surface, the on-chip ESR is a convenient tool for this investigation, see Fig.~\ref{fig:Setup}(d).

Figure~\ref{fig:Setup}(b) shows a picture of the YSO crystal magnetically coupled to an array of superconducting lumped element (LE) resonators~\cite{Probst_PRL13}. In our work, we investigate two samples (YSO2 and YSO4), which have 5 and 4 implanted areas, respectively. Each implanted area covers one specific LE resonator. An optical micrograph of a single resonator is displayed in Fig.~\ref{fig:Setup}(c). The LE resonators covered by the Er:YSO crystal possess loaded quality factors of $Q_l\sim 650$. The resonance frequencies of the LE resonators cover the frequency range from 4 to 5 GHz.

Figure~\ref{fig:Setup}(d) presents a schematic cross-section of the coupling area. The oscillating magnetic field of the microwave penetrates several microns into the crystal, see Ref.~\cite{Probst_PRL13} for a simulation of the AC field. A DC magnetic field is applied along the surface of the SC chip. The experiments were carried out inside a BlueFors LD-250 dilution fridge at a base temperature of 25\,mK. The on-chip ESR was performed inside a superconducting solenoid coil with a maximum magnetic field of 370\,mT.

\begin{figure}[ht!]
\includegraphics[width=0.8\columnwidth]{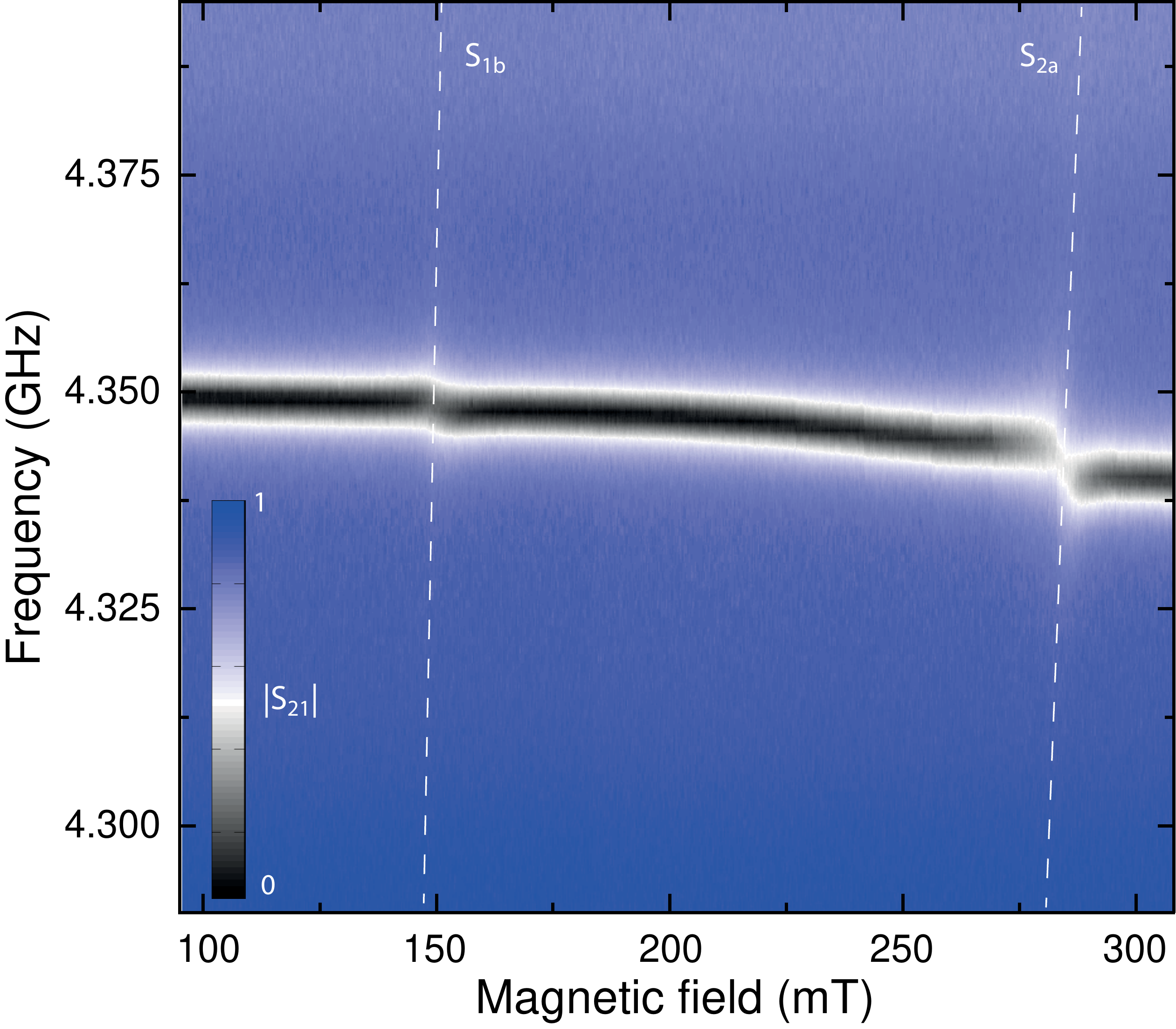}
\caption{(Color online) ESR transmission spectrum of the YSO4 crystal, area \#3 coupled to one of the LE resonators. Two dispersive shifts of the resonator are visible in the spectrum due to the magnetic coupling to the electronic spin transitions $S_{1b}$ and $S_{2a}$ of the erbium ions~\cite{Probst_PRL13}.}
\label{fig:ESRspec}
\end{figure}

Figure~\ref{fig:ESRspec} shows the on-chip ESR spectrum of the YSO4 sample area \#3. The color plot displays the transmitted amplitude $\left|S_{21}\right|$ as a function of the magnetic field and the probe frequency. The resonator shows up as a black line which is distorted by two dispersive shifts induced by weakly coupled Er spins~\cite{Shuster2010, Bushev2011}. From the spectrum we extract the collective coupling strength $v$ and inhomogeneous spin linewidth $\Gamma^*_2$. In the following, we will focus on the high field transition.

The collective coupling strength of a resonator to a spin ensemble of constant spin density $n$ is given by $v_N = \tilde{\textrm{g}} \mu_B / \hbar \sqrt{\mu_0 \hbar \omega_r n \xi /4}$, see Ref.~\cite{Bushev2011}, where $\omega_r$ is the frequency of the resonator, $\xi$ the filling factor and $\tilde{\textrm{g}}$ the effective g-factor. The collective coupling strength of a spin ensemble of size $N$ to the resonator is related to the coupling strength per single spin $v_1$ by $v_N=v_1\sqrt{N}$. In the limit of weak coupling, where the inhomogeneous linewidth of the spin ensemble $\Gamma^*_2$ exceeds the coupling strength $v$, the HWHM linewidth of the superconducting microwave resonator $\kappa$ is given by \cite{Bushev2011}
\begin{equation}
\kappa(B) = \kappa_0 + \frac{v_N^2 \Gamma^*_2}{\left( \omega_r - \gamma B_z \right)^2 + {\Gamma^*_2}^2} \,.
\label{eq:k0dep}
\end{equation}
Here, the spin frequency is given by the Zeeman shift $\omega_S=\gamma B_z$ with spin tuning rate $\gamma=\mu_B/\textrm{g}_\text{dc}\hbar$. Er spins in YSO possess a large magnetic anisotropy and it was shown that maximum coupling strength is obtained for small DC g-factors and large AC g-factors, respectively\cite{Probst_PRL13}. The DC g-factors for the high field transition of our samples are 1.09 (YSO4) and $1.03$ (YSO2), respectively, and the orientation of the crystals is shown in Fig.~\ref{fig:Setup}(a).

\begin{figure}[ht!]
\includegraphics[width=1\columnwidth]{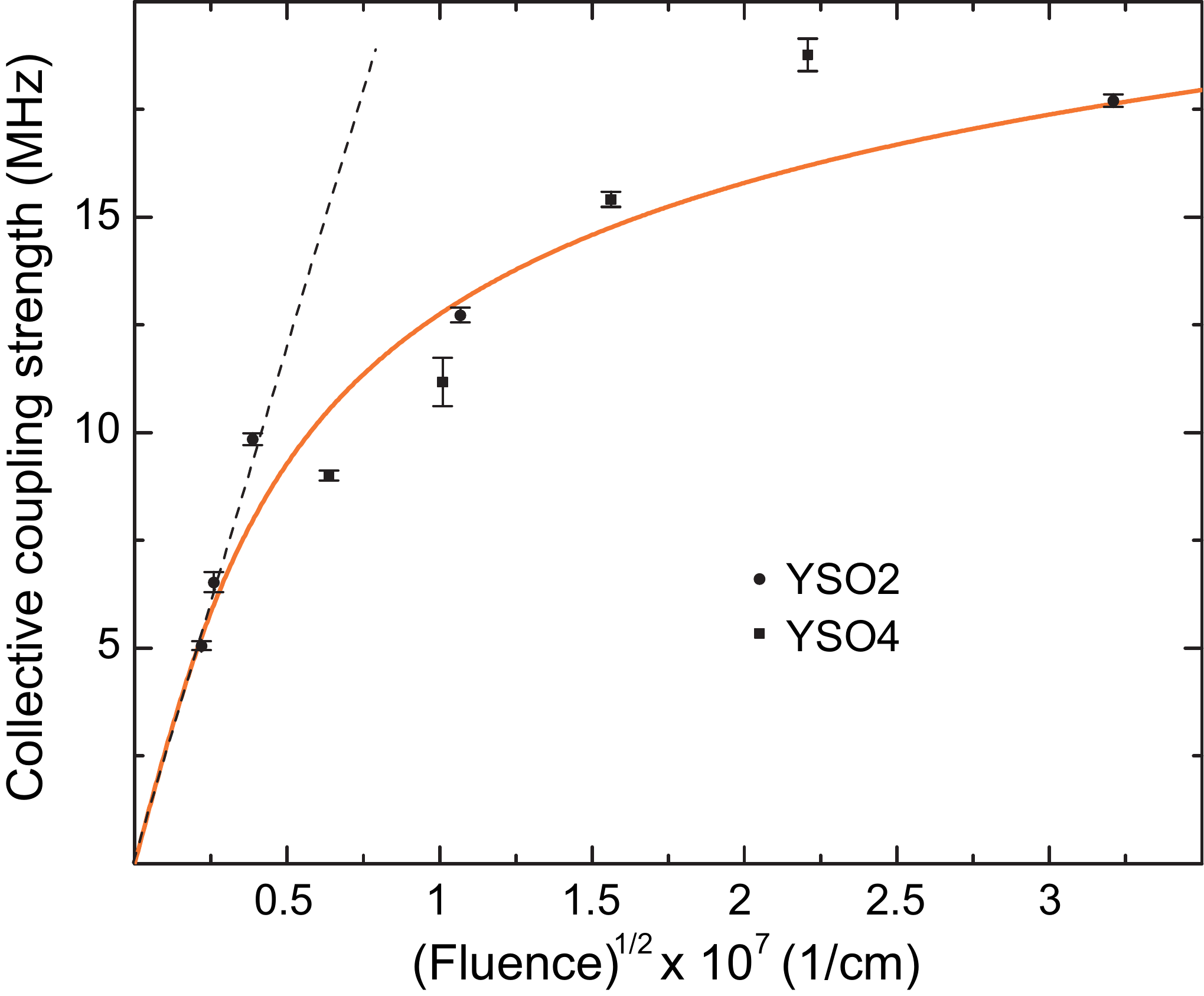}
\caption{(Color online) Collective coupling strength $v_N$ as a function of the square-root of the fluence $F$. The presented data do not follow the expected dependence $v_N\propto \sqrt{F}$ (dashed line). The solid line shows the fit to an empirical model, see text for details.}
\label{fig:dopingdep}
\end{figure}

We analyze the spectra of all samples by extracting the collective coupling strength $v_N$ and inhomogeneous spin linewidth $\Gamma^*_2$ as a function of the implanted fluence. The coupling strength varies between 5\,MHz and 18.7\,MHz, while the linewidths range from 50\,MHz to 130\,MHz. Figure~\ref{fig:dopingdep} displays the extracted coupling strengths versus the square root of the fluences. If the implantation process had the same efficiency for all fluences, one would expect the data points to follow a straight line because $v_N \propto \sqrt{F}$. However, the coupling strength is not proportional to the square root of the incident fluence $F$ and the deviation increases for larger fluences.

In order to interpret this result, it is important to emphasize that the entire sample preparation process consists of FIB irradiation and subsequent annealing in Ar atmosphere. Both processes can provide a contribution to the reduced implantation efficiency. Two samples, which were only treated with RTA, show no ESR response. We found that long term annealing ($\sim$ 1.5-2 hours) is crucial in order to detect the erbium ions with on-chip ESR. However, a large inhomogeneous spin linewidth remains and does not show a significant dependence on the fluence.

There are two major contributions to the inhomogeneous spin linewidth: Dipole-dipole interaction and local distortions of the crystal field which modulates the effective g-factor. Using Refs.~\cite{Morton2011, Bhattacharyya2008}, we can estimate an upper limit for the contribution to the inhomogeneous broadening due to dipole-dipole interaction. Assuming the ion distribution calculated by our Monte Carlo simulation, $\Gamma^*_2/2\pi$ varies from tens of kHz for the lowest fluence to approximately 80~MHz for the sample with the largest fluence. 
\begin{figure}[ht!]
\includegraphics[width=1\columnwidth]{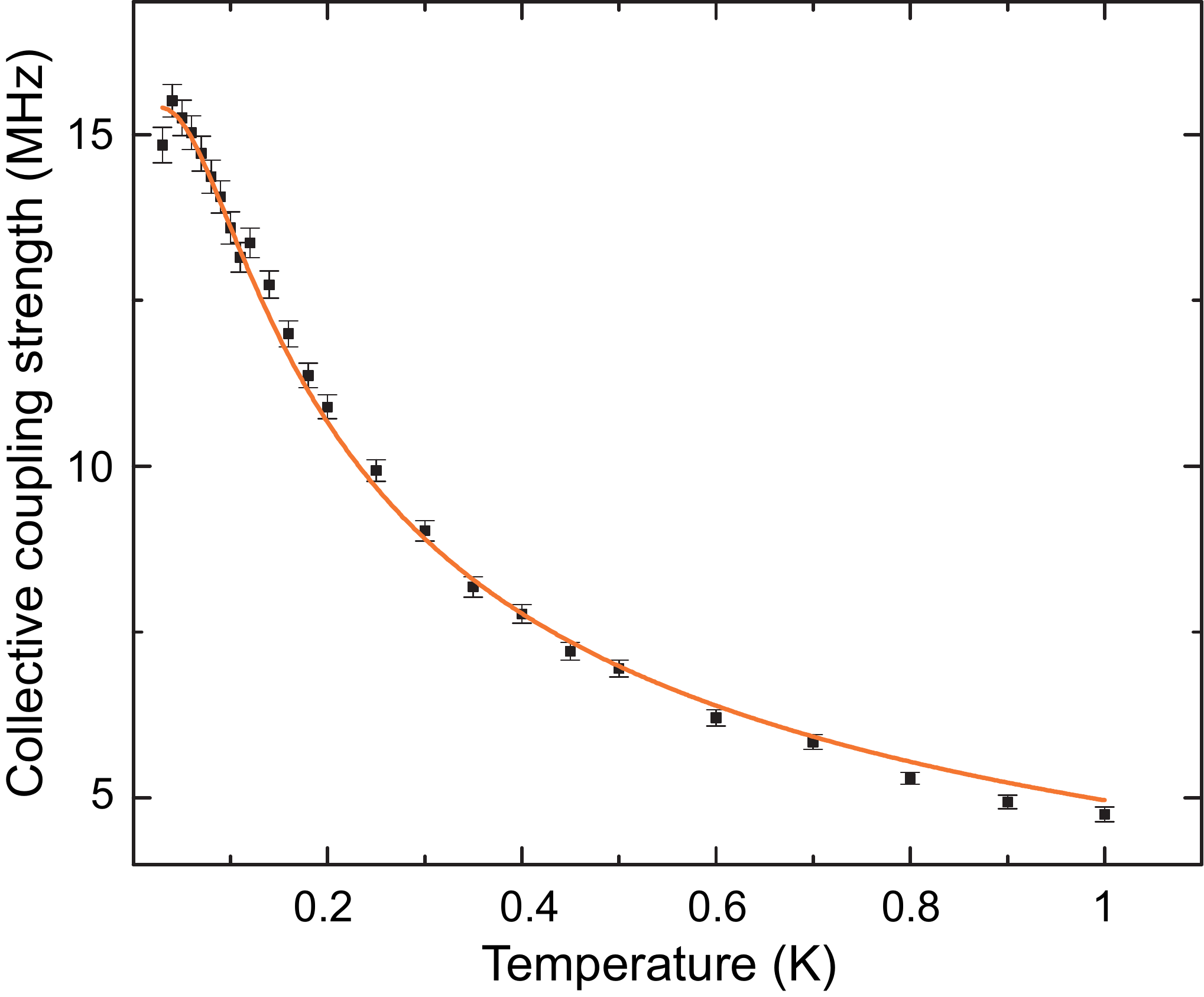}
\caption{(Color online) Temperature dependence of the collective coupling strength $v_N$ of sample YSO4\#3. The coupling shows nice agreement with the standard theory of paramagnetic ions.}
\label{fig:tempdep}
\end{figure}
Since the linewidth always resides at a high level, we conclude that the distorted crystal field dominates the linewidth. The details of the implantation and annealing processes are complex and a more detailed study would be needed to model the entire process. 

Since we do not know exactly the distribution of the ions in the crystal and the distortion of the lattice, we use the known implantation process parameter fluence to motivate a simplified empirical model. From all erbium ions per area $F$, which penetrate the sample, only a subset $D\leq F$ will finally be a in an appropriate configuration. Therefore, merely these ions contribute to the ESR signal and $v_N \propto \sqrt{D}$. We assume that the implantation yield (including post-annealing) is exponentially reduced by the amount of ions per implanted area $D$ (dose), $\dot{D} = k\, \exp (-D/D_c )$.
Here,  $k=I/S$ is the intensity of the FIB and $D_c$ denotes a critical dose where local interaction effects in the crystal start to dominate over the enhanced amount of erbium ions. The total number of incident ions per area is given by the fluence $F=kt$, where $t$ denotes the total implantation time for a given area $S$. The effective implanted dose $D$ with respect to the fluence is given by
\begin{equation}
D(F) =\ln\left(1+\frac{F}{D_c}\right) D_c \,,
\label{eq:fluencecorr}
\end{equation}
thus, $v_N \propto \sqrt{D(F)}$. The fit to the experimental points is presented in Fig.~\ref{fig:dopingdep} and yields $D_c\approx 10^{13}$~cm$^{-2}$. Only in the limit of low fluence $F \ll D_c$, the collective coupling is proportional to the total number of implanted ions.

\begin{table*}[!ht]
\begin{ruledtabular}
\begin{tabular}{c|cccccccccc}
	crystal ID&\multicolumn{5}{c}{YSO2} & \multicolumn{4}{c}{YSO4} & bulk\\
  implanted area & \#1 & \#2 & \#3 & \#4 & \#5 & \#1 & \#2 & \#3 & \#4 &  Er:YSO\cite{Probst_PRL13}\\ \hline
    $v$/2$\pi$ (MHz) & 5.06 & 6.53 & 9.85 & 12.73 & 17.71 & 9.01 & 11.18 & 15.41 & 18.77 & 34\\
    $\Gamma^*_2$/2$\pi$ (MHz) &  68.1 & 107 & 105 & 105 & 112 & 48.6 & 136 & 69.2 & 93.1 & 12 \\
    $F$ ($10^{12}$cm$^{-2}$) & 4.80 & 6.73 & 15.0 & 114 & 1030 & 40.6 & 102 & 244&487 & -- \\
		$\tilde{c}_\text{eff}$$^\ddagger$ (ppm) & 35 & 46 & 81 & 225 & 414 & 145 & 215 & 288 & 348 & 200 \\
		$\tilde{N}_\text{eff}$ ($\times 10^{9}$) & 14.1 & 18.5 & 32.9 & 90.8 & 167 & 58.4 & 86.8 & 116 & 141 & $\sim 10^3$ \\
		$\bar{v}_1/2\pi$ (Hz) & 85 & 96 & 109 & 85 & 87 & 75 & 76 & 90 & 100 & 76 \\
    $T^a_\text{SLR}$ (s) & $^\dagger$ & $^\dagger$ & 1.13 & 1.10 & 1.19 & 1.83&2.91&1.9&(2)$^*$ & 4.3 \\
		$T^b_\text{SLR}$ (s) & $^\dagger$ & $^\dagger$ & 7.36 & 8.16 & 7.49 & 12.2 & 15.6  & 11.9& (14)$^\sharp$ & --\\ 
		annealing param.& \multicolumn{5}{c}{2~h at $1200^\circ$C in Ar atmosphere} & \multicolumn{4}{c}{1.5~h at $1200^\circ$C in Ar atmosphere} & --
  \end{tabular}
	\end{ruledtabular}
	\begin{flushleft}
	\scriptsize$^\ddagger$assuming an average 60~nm thick layer, $^*$uncertainty $\approx 30$\%, $^\sharp$uncertainty $\approx 25$\%, $^\dagger$SLR signal too weak
	\end{flushleft}
  \caption{Comparison of coupling strength $v$, inhomogeneous spin linewidth $\Gamma^*_2$, fluence $F$, effective local concentration $\tilde{c}_\text{eff}$, effective number of spins $\tilde{N}_\text{eff}$ coupled to the transition, the average single spin coupling strength $\bar{v}_1$ and spin lattice relaxation times $T^a_\text{SLR}$, $T^b_\text{SLR}$ of all samples and a doped as-grown crystal from an earlier publication\cite{Probst_PRL13}.}
	\label{tab:specs}
\end{table*}

In order to check whether the implanted ions form a paramagnetic spin ensemble, we studied the temperature dependence for sample YSO4 area \#3.
Figure \ref{fig:tempdep} shows the measured temperature dependence of the collective coupling strength $v_N$, which nicely follows the theory of paramagnetic crystals
\begin{equation}
v(T) = v_0 \sqrt{\tanh\left( \frac{\hbar \omega}{2 k_B T}\right)}\,.
\label{eq:tempdep}
\end{equation}
Thus, the implanted ions can be modelled as a system of the independent spins~\cite{Bushev2011}.

In contrast to the coupling strength the inhomogeneous spin linewidth $\Gamma^*_2$ is about $2\pi\times$70\,MHz and stays constant in the temperature range from 30\,mK up to 1\,K. Since crystals grown by the Czochralski method are known to have a much smaller linewidth (12 MHz for 200 ppm doping concentration) at low temperatures~\cite{Probst_PRL13}, we assume, that the distorted crystal field dominates the inhomogeneous broadening over dynamic effects, in this temperature range.

In the limit of weak coupling where $v_N\ll \Gamma^*_2$ and the cooperativity $C=v_N^2/\kappa \Gamma^*_2 < 1$, we can study the spin relaxation dynamics directly. Here, we employ a technique similar to Ref.~\cite{Probst_PRL13}. In contrast to that reference, our weakly coupled cavity is equivalent to a long transmission line. The decay rate of the resonator tuned in resonance with the spins is $\kappa=\kappa_0+v_N^2/\Gamma^*_2$. The amplitude of the depth of the resonator dip is proportional to $1/\kappa= [\kappa_0\left(1+v_N^2/\kappa_0\Gamma^*_2\right)]^{-1} \approx 1/\kappa_0-v_N^2/\kappa_0^2\Gamma^*_2$. The second term in that equation is proportional to the number of participating spins given by the population difference $v_N^2 \propto N_2-N_1$.

An initial strong microwave pulse saturates the spin ensemble and the depth of the resonator dip increases significantly. The resonator is then continuously probed at low power in order to observe the decrease of the resonator dip while the spin ensemble is relaxing back to its equilibrium position. We observe a double exponential decay with two time scales $T^a_{\text{SLR}}\approx 1\,s$ and $T^b_{\text{SLR}} \approx 10\,s$. This suggests that the energy relaxation is dominated by a direct process on short timescales plus an indirect process on larger timescales. Typically, the spin-lattice relaxation time $T_1$ at low temperatures ($\hbar \omega \gg k_B T$) is dominated by a direct process $1/T_1 \propto (\hbar \omega)^5 \coth(\hbar \omega/2k_B T)$, see Ref.~\cite{AbragamESR} chapter 10.

Table~\ref{tab:specs} summarizes the results of our on-chip ESR investigations of implanted Er:YSO samples and compares them to a doped as-grown Er:YSO crystal. We note here, that the implanted crystals have a similar orientation as the bulk doped crystal, see also Ref.~\cite{Probst_PRL13}. The effective local concentration and number of spins are calculated from $F$ and using Eq.~(\ref{eq:fluencecorr}). All implanted samples couple weakly to the LE resonators because $v\ll \Gamma^*_2$. The linewidth $\Gamma^*_2$ does not show a significant dependence on the coupling strength or fluence. The value of the calculated average single-spin coupling is in accordance with our estimation of $\bar{v}_1/2\pi\sim 70$~Hz using the inductance of the LE resonator \cite{Wuensch2011} and the simulated magnetic field \cite{Probst_PRL13} with an AC g-factor of 15. We detect no significant difference between the YSO2 and the YSO4 samples.

To conclude, erbium ions were successfully implanted by focused ion beam irradiation of a Y$_2$SiO$_5$ substrate. The ions possessed an energy of 300~keV and penetrated up to 100\,nm inside the substrate. A subsequent annealing of the samples in argon atmosphere at 1200~$^\circ$C for 1.5-2 hours turned out to be crucial. The implanted spin ensemble was characterized by on-chip ESR spectroscopy at 20 mK and at the single microwave photon limit. The collective coupling strengths of the implanted spins vary from 5 to 18.7~MHz, and exceed the typical dissipation rates of SC circuits. The temperature dependence of the coupling strength shows paramagnetic behavior. The inhomogeneous spin linewidth is 5 to 10 times larger compared to a bulk doped Er:YSO crystal grown by the Czochralski method. In order to reduce the spin linewidth, further investigations of the annealing process are required. We believe that implantation into heated substrates will be crucial for an optimal process. Our work paves the way towards the local integration of erbium spins in SC quantum circuits. This concept can be employed for the reversible conversion of microwave and telecom C-band photons by combining microwave and optical waveguides on the same crystal~\cite{Pernice2013, Morigi2014}.

S.~P.~acknowledges financial support by the LGF of Baden-W\"{u}rttemberg. We thank Ch.~Hintze and M.~Drescher for the ESR characterization of undoped and grown YSO crystals. This work was supported in part by the DFG, the BMBF program "Quantum communications" through the project QUIMP and the Ministry of Education and Science of the Russian Federation under contract no.~11.G34.31.0062. N.~K.~and A.~D.~W.~acknowledge gratefully support of Mercur Pr-2013-0001, the DFH/UFA  CDFA-05-06 and RUB Research School.

\bibliographystyle{apsrev4-1}
\bibliography{ESRimplanted}

\begin{thebibliography}{34}%
\makeatletter
\providecommand \@ifxundefined [1]{%
 \@ifx{#1\undefined}
}%
\providecommand \@ifnum [1]{%
 \ifnum #1\expandafter \@firstoftwo
 \else \expandafter \@secondoftwo
 \fi
}%
\providecommand \@ifx [1]{%
 \ifx #1\expandafter \@firstoftwo
 \else \expandafter \@secondoftwo
 \fi
}%
\providecommand \natexlab [1]{#1}%
\providecommand \enquote  [1]{``#1''}%
\providecommand \bibnamefont  [1]{#1}%
\providecommand \bibfnamefont [1]{#1}%
\providecommand \citenamefont [1]{#1}%
\providecommand \href@noop [0]{\@secondoftwo}%
\providecommand \href [0]{\begingroup \@sanitize@url \@href}%
\providecommand \@href[1]{\@@startlink{#1}\@@href}%
\providecommand \@@href[1]{\endgroup#1\@@endlink}%
\providecommand \@sanitize@url [0]{\catcode `\\12\catcode `\$12\catcode
  `\&12\catcode `\#12\catcode `\^12\catcode `\_12\catcode `\%12\relax}%
\providecommand \@@startlink[1]{}%
\providecommand \@@endlink[0]{}%
\providecommand \url  [0]{\begingroup\@sanitize@url \@url }%
\providecommand \@url [1]{\endgroup\@href {#1}{\urlprefix }}%
\providecommand \urlprefix  [0]{URL }%
\providecommand \Eprint [0]{\href }%
\providecommand \doibase [0]{http://dx.doi.org/}%
\providecommand \selectlanguage [0]{\@gobble}%
\providecommand \bibinfo  [0]{\@secondoftwo}%
\providecommand \bibfield  [0]{\@secondoftwo}%
\providecommand \translation [1]{[#1]}%
\providecommand \BibitemOpen [0]{}%
\providecommand \bibitemStop [0]{}%
\providecommand \bibitemNoStop [0]{.\EOS\space}%
\providecommand \EOS [0]{\spacefactor3000\relax}%
\providecommand \BibitemShut  [1]{\csname bibitem#1\endcsname}%
\let\auto@bib@innerbib\@empty
\bibitem [{\citenamefont {Wallquist}\ \emph {et~al.}(2009)\citenamefont
  {Wallquist}, \citenamefont {Hammerer}, \citenamefont {Rabl}, \citenamefont
  {Lukin},\ and\ \citenamefont {Zoller}}]{Zoller2009}%
  \BibitemOpen
  \bibfield  {author} {\bibinfo {author} {\bibfnamefont {M.}~\bibnamefont
  {Wallquist}}, \bibinfo {author} {\bibfnamefont {K.}~\bibnamefont {Hammerer}},
  \bibinfo {author} {\bibfnamefont {P.}~\bibnamefont {Rabl}}, \bibinfo {author}
  {\bibfnamefont {M.~D.}\ \bibnamefont {Lukin}}, \ and\ \bibinfo {author}
  {\bibfnamefont {P.}~\bibnamefont {Zoller}},\ }\href@noop {} {\bibfield
  {journal} {\bibinfo  {journal} {Phys. Scr.}\ }\textbf {\bibinfo {volume}
  {T137}},\ \bibinfo {pages} {014001} (\bibinfo {year} {2009})}\BibitemShut
  {NoStop}%
\bibitem [{\citenamefont {Xiang}\ \emph {et~al.}(2013)\citenamefont {Xiang},
  \citenamefont {Asshab}, \citenamefont {You},\ and\ \citenamefont
  {Nori}}]{Nori2013}%
  \BibitemOpen
  \bibfield  {author} {\bibinfo {author} {\bibfnamefont {Z.}~\bibnamefont
  {Xiang}}, \bibinfo {author} {\bibfnamefont {S.}~\bibnamefont {Asshab}},
  \bibinfo {author} {\bibfnamefont {J.}~\bibnamefont {You}}, \ and\ \bibinfo
  {author} {\bibfnamefont {F.}~\bibnamefont {Nori}},\ }\href@noop {} {\bibfield
   {journal} {\bibinfo  {journal} {Rev.~Mod.~Phys.}\ }\textbf {\bibinfo
  {volume} {85}},\ \bibinfo {pages} {623} (\bibinfo {year} {2013})}\BibitemShut
  {NoStop}%
\bibitem [{\citenamefont {Clarke}\ and\ \citenamefont
  {Wilhelm}(2008)}]{Clarke2008}%
  \BibitemOpen
  \bibfield  {author} {\bibinfo {author} {\bibfnamefont {J.}~\bibnamefont
  {Clarke}}\ and\ \bibinfo {author} {\bibfnamefont {F.}~\bibnamefont
  {Wilhelm}},\ }\href@noop {} {\bibfield  {journal} {\bibinfo  {journal}
  {Nature}\ }\textbf {\bibinfo {volume} {453}},\ \bibinfo {pages} {1031}
  (\bibinfo {year} {2008})}\BibitemShut {NoStop}%
\bibitem [{\citenamefont {Bialczak}\ \emph {et~al.}(2010)\citenamefont
  {Bialczak}, \citenamefont {Ansmann}, \citenamefont {Hofheinz}, \citenamefont
  {Lucero}, \citenamefont {Neeley}, \citenamefont {O'Connell}, \citenamefont
  {Sank}, \citenamefont {Wang}, \citenamefont {Wenner}, \citenamefont
  {Steffen}, \citenamefont {Cleland},\ and\ \citenamefont
  {Martinis}}]{Martinis2010}%
  \BibitemOpen
  \bibfield  {author} {\bibinfo {author} {\bibfnamefont {R.~C.}\ \bibnamefont
  {Bialczak}}, \bibinfo {author} {\bibfnamefont {M.}~\bibnamefont {Ansmann}},
  \bibinfo {author} {\bibfnamefont {M.}~\bibnamefont {Hofheinz}}, \bibinfo
  {author} {\bibfnamefont {E.}~\bibnamefont {Lucero}}, \bibinfo {author}
  {\bibfnamefont {M.}~\bibnamefont {Neeley}}, \bibinfo {author} {\bibfnamefont
  {A.~D.}\ \bibnamefont {O'Connell}}, \bibinfo {author} {\bibfnamefont
  {D.}~\bibnamefont {Sank}}, \bibinfo {author} {\bibfnamefont {H.}~\bibnamefont
  {Wang}}, \bibinfo {author} {\bibfnamefont {J.}~\bibnamefont {Wenner}},
  \bibinfo {author} {\bibfnamefont {M.}~\bibnamefont {Steffen}}, \bibinfo
  {author} {\bibfnamefont {A.~N.}\ \bibnamefont {Cleland}}, \ and\ \bibinfo
  {author} {\bibfnamefont {J.~M.}\ \bibnamefont {Martinis}},\ }\href@noop {}
  {\bibfield  {journal} {\bibinfo  {journal} {Nat.~Phys}\ }\textbf {\bibinfo
  {volume} {6}},\ \bibinfo {pages} {409} (\bibinfo {year} {2010})}\BibitemShut
  {NoStop}%
\bibitem [{\citenamefont {Tyryshkin}\ \emph {et~al.}(2011)\citenamefont
  {Tyryshkin}, \citenamefont {Tojo}, \citenamefont {Morton}, \citenamefont
  {Riemann}, \citenamefont {Abrosimov}, \citenamefont {Becker}, \citenamefont
  {Pohl}, \citenamefont {Schenkel}, \citenamefont {Thewalt}, \citenamefont
  {Itoh},\ and\ \citenamefont {Lyon}}]{Tyryshkin2011}%
  \BibitemOpen
  \bibfield  {author} {\bibinfo {author} {\bibfnamefont {A.~M.}\ \bibnamefont
  {Tyryshkin}}, \bibinfo {author} {\bibfnamefont {S.}~\bibnamefont {Tojo}},
  \bibinfo {author} {\bibfnamefont {J.~J.~L.}\ \bibnamefont {Morton}}, \bibinfo
  {author} {\bibfnamefont {H.}~\bibnamefont {Riemann}}, \bibinfo {author}
  {\bibfnamefont {N.~V.}\ \bibnamefont {Abrosimov}}, \bibinfo {author}
  {\bibfnamefont {P.}~\bibnamefont {Becker}}, \bibinfo {author} {\bibfnamefont
  {H.-J.}\ \bibnamefont {Pohl}}, \bibinfo {author} {\bibfnamefont
  {T.}~\bibnamefont {Schenkel}}, \bibinfo {author} {\bibfnamefont {M.~L.~W.}\
  \bibnamefont {Thewalt}}, \bibinfo {author} {\bibfnamefont {K.~M.}\
  \bibnamefont {Itoh}}, \ and\ \bibinfo {author} {\bibfnamefont {S.~A.}\
  \bibnamefont {Lyon}},\ }\href@noop {} {\bibfield  {journal} {\bibinfo
  {journal} {Nat.~Mat.}\ }\textbf {\bibinfo {volume} {11}},\ \bibinfo {pages}
  {143} (\bibinfo {year} {2011})}\BibitemShut {NoStop}%
\bibitem [{\citenamefont {Kubo}\ \emph {et~al.}(2011)\citenamefont {Kubo},
  \citenamefont {Grezes}, \citenamefont {Dewes}, \citenamefont {Umeda},
  \citenamefont {Isoya}, \citenamefont {Morishita}, \citenamefont {Abe},
  \citenamefont {Onoda}, \citenamefont {Ohshima}, \citenamefont {Jacques},
  \citenamefont {Dr\'{e}au}, \citenamefont {Roch}, \citenamefont {Diniz},
  \citenamefont {Auffeves}, \citenamefont {Vion}, \citenamefont {Esteve},\ and\
  \citenamefont {Bertet}}]{Bertet2011}%
  \BibitemOpen
  \bibfield  {author} {\bibinfo {author} {\bibfnamefont {Y.}~\bibnamefont
  {Kubo}}, \bibinfo {author} {\bibfnamefont {C.}~\bibnamefont {Grezes}},
  \bibinfo {author} {\bibfnamefont {A.}~\bibnamefont {Dewes}}, \bibinfo
  {author} {\bibfnamefont {T.}~\bibnamefont {Umeda}}, \bibinfo {author}
  {\bibfnamefont {J.}~\bibnamefont {Isoya}}, \bibinfo {author} {\bibfnamefont
  {H.~S.~N.}\ \bibnamefont {Morishita}}, \bibinfo {author} {\bibfnamefont
  {H.}~\bibnamefont {Abe}}, \bibinfo {author} {\bibfnamefont {S.}~\bibnamefont
  {Onoda}}, \bibinfo {author} {\bibfnamefont {T.}~\bibnamefont {Ohshima}},
  \bibinfo {author} {\bibfnamefont {V.}~\bibnamefont {Jacques}}, \bibinfo
  {author} {\bibfnamefont {A.}~\bibnamefont {Dr\'{e}au}}, \bibinfo {author}
  {\bibfnamefont {J.-F.}\ \bibnamefont {Roch}}, \bibinfo {author}
  {\bibfnamefont {I.}~\bibnamefont {Diniz}}, \bibinfo {author} {\bibfnamefont
  {A.}~\bibnamefont {Auffeves}}, \bibinfo {author} {\bibfnamefont
  {D.}~\bibnamefont {Vion}}, \bibinfo {author} {\bibfnamefont {D.}~\bibnamefont
  {Esteve}}, \ and\ \bibinfo {author} {\bibfnamefont {P.}~\bibnamefont
  {Bertet}},\ }\href@noop {} {\bibfield  {journal} {\bibinfo  {journal} {Phys.
  Rev. Lett.}\ }\textbf {\bibinfo {volume} {107}},\ \bibinfo {pages} {220501}
  (\bibinfo {year} {2011})}\BibitemShut {NoStop}%
\bibitem [{\citenamefont {Lauritzen}\ \emph {et~al.}(2010)\citenamefont
  {Lauritzen}, \citenamefont {Min\'{a}\v{r}}, \citenamefont {de~Riedmatten},
  \citenamefont {Afzelius}, \citenamefont {Sangouard}, \citenamefont {Simon},\
  and\ \citenamefont {Gisin}}]{Gisin2010}%
  \BibitemOpen
  \bibfield  {author} {\bibinfo {author} {\bibfnamefont {B.}~\bibnamefont
  {Lauritzen}}, \bibinfo {author} {\bibfnamefont {J.}~\bibnamefont
  {Min\'{a}\v{r}}}, \bibinfo {author} {\bibfnamefont {H.}~\bibnamefont
  {de~Riedmatten}}, \bibinfo {author} {\bibfnamefont {M.}~\bibnamefont
  {Afzelius}}, \bibinfo {author} {\bibfnamefont {N.}~\bibnamefont {Sangouard}},
  \bibinfo {author} {\bibfnamefont {C.}~\bibnamefont {Simon}}, \ and\ \bibinfo
  {author} {\bibfnamefont {N.}~\bibnamefont {Gisin}},\ }\href@noop {}
  {\bibfield  {journal} {\bibinfo  {journal} {Phys. Rev. Lett.}\ }\textbf
  {\bibinfo {volume} {104}},\ \bibinfo {pages} {080502} (\bibinfo {year}
  {2010})}\BibitemShut {NoStop}%
\bibitem [{\citenamefont {Hedges}\ \emph {et~al.}(2010)\citenamefont {Hedges},
  \citenamefont {Longdell}, \citenamefont {Li},\ and\ \citenamefont
  {Sellars}}]{Sellars2010}%
  \BibitemOpen
  \bibfield  {author} {\bibinfo {author} {\bibfnamefont {M.}~\bibnamefont
  {Hedges}}, \bibinfo {author} {\bibfnamefont {J.}~\bibnamefont {Longdell}},
  \bibinfo {author} {\bibfnamefont {Y.}~\bibnamefont {Li}}, \ and\ \bibinfo
  {author} {\bibfnamefont {M.}~\bibnamefont {Sellars}},\ }\href@noop {}
  {\bibfield  {journal} {\bibinfo  {journal} {Nature}\ }\textbf {\bibinfo
  {volume} {465}},\ \bibinfo {pages} {1053} (\bibinfo {year}
  {2010})}\BibitemShut {NoStop}%
\bibitem [{\citenamefont {Clausen}\ \emph {et~al.}(2011)\citenamefont
  {Clausen}, \citenamefont {Usmani}, \citenamefont {Bussi\`{e}res},
  \citenamefont {Sangouard}, \citenamefont {Afzelius}, \citenamefont
  {de~Riedmatten},\ and\ \citenamefont {Gisin}}]{Gisin2011}%
  \BibitemOpen
  \bibfield  {author} {\bibinfo {author} {\bibfnamefont {C.}~\bibnamefont
  {Clausen}}, \bibinfo {author} {\bibfnamefont {I.}~\bibnamefont {Usmani}},
  \bibinfo {author} {\bibfnamefont {F.}~\bibnamefont {Bussi\`{e}res}}, \bibinfo
  {author} {\bibfnamefont {N.}~\bibnamefont {Sangouard}}, \bibinfo {author}
  {\bibfnamefont {M.}~\bibnamefont {Afzelius}}, \bibinfo {author}
  {\bibfnamefont {H.}~\bibnamefont {de~Riedmatten}}, \ and\ \bibinfo {author}
  {\bibfnamefont {N.}~\bibnamefont {Gisin}},\ }\href@noop {} {\bibfield
  {journal} {\bibinfo  {journal} {Nature}\ }\textbf {\bibinfo {volume} {469}},\
  \bibinfo {pages} {508} (\bibinfo {year} {2011})}\BibitemShut {NoStop}%
\bibitem [{\citenamefont {Probst}\ \emph {et~al.}(2013)\citenamefont {Probst},
  \citenamefont {Rotzinger}, \citenamefont {W\"{u}nsch}, \citenamefont {Jung},
  \citenamefont {Jerger}, \citenamefont {Siegel}, \citenamefont {Ustinov},\
  and\ \citenamefont {Bushev}}]{Probst_PRL13}%
  \BibitemOpen
  \bibfield  {author} {\bibinfo {author} {\bibfnamefont {S.}~\bibnamefont
  {Probst}}, \bibinfo {author} {\bibfnamefont {H.}~\bibnamefont {Rotzinger}},
  \bibinfo {author} {\bibfnamefont {S.}~\bibnamefont {W\"{u}nsch}}, \bibinfo
  {author} {\bibfnamefont {P.}~\bibnamefont {Jung}}, \bibinfo {author}
  {\bibfnamefont {M.}~\bibnamefont {Jerger}}, \bibinfo {author} {\bibfnamefont
  {M.}~\bibnamefont {Siegel}}, \bibinfo {author} {\bibfnamefont {A.~V.}\
  \bibnamefont {Ustinov}}, \ and\ \bibinfo {author} {\bibfnamefont {P.~A.}\
  \bibnamefont {Bushev}},\ }\href {\doibase 10.1103/physrevlett.110.157001}
  {\bibfield  {journal} {\bibinfo  {journal} {Phys. Rev. Lett.}\ }\textbf
  {\bibinfo {volume} {110}},\ \bibinfo {pages} {157001} (\bibinfo {year}
  {2013})}\BibitemShut {NoStop}%
\bibitem [{\citenamefont {Usmani}\ \emph {et~al.}(2012)\citenamefont {Usmani},
  \citenamefont {Clausen}, \citenamefont {Bussieres}, \citenamefont
  {Sangouard}, \citenamefont {Afzelius},\ and\ \citenamefont
  {Gisin}}]{Usmani2012}%
  \BibitemOpen
  \bibfield  {author} {\bibinfo {author} {\bibfnamefont {I.}~\bibnamefont
  {Usmani}}, \bibinfo {author} {\bibfnamefont {C.}~\bibnamefont {Clausen}},
  \bibinfo {author} {\bibfnamefont {F.}~\bibnamefont {Bussieres}}, \bibinfo
  {author} {\bibfnamefont {N.}~\bibnamefont {Sangouard}}, \bibinfo {author}
  {\bibfnamefont {M.}~\bibnamefont {Afzelius}}, \ and\ \bibinfo {author}
  {\bibfnamefont {N.}~\bibnamefont {Gisin}},\ }\href
  {http://dx.doi.org/10.1038/nphoton.2012.34} {\bibfield  {journal} {\bibinfo
  {journal} {Nat Photon}\ }\textbf {\bibinfo {volume} {6}},\ \bibinfo {pages}
  {234} (\bibinfo {year} {2012})}\BibitemShut {NoStop}%
\bibitem [{\citenamefont {Sabooni}\ \emph {et~al.}(2013)\citenamefont
  {Sabooni}, \citenamefont {Li}, \citenamefont {Kr\"oll},\ and\ \citenamefont
  {Rippe}}]{Kroell2013}%
  \BibitemOpen
  \bibfield  {author} {\bibinfo {author} {\bibfnamefont {M.}~\bibnamefont
  {Sabooni}}, \bibinfo {author} {\bibfnamefont {Q.}~\bibnamefont {Li}},
  \bibinfo {author} {\bibfnamefont {S.}~\bibnamefont {Kr\"oll}}, \ and\
  \bibinfo {author} {\bibfnamefont {L.}~\bibnamefont {Rippe}},\ }\href
  {\doibase 10.1103/PhysRevLett.110.133604} {\bibfield  {journal} {\bibinfo
  {journal} {Phys. Rev. Lett.}\ }\textbf {\bibinfo {volume} {110}},\ \bibinfo
  {pages} {133604} (\bibinfo {year} {2013})}\BibitemShut {NoStop}%
\bibitem [{\citenamefont {Riel\"ander}\ \emph {et~al.}(2014)\citenamefont
  {Riel\"ander}, \citenamefont {Kutluer}, \citenamefont {Ledingham},
  \citenamefont {G\"undo\ifmmode~\breve{g}\else \u{g}\fi{}an}, \citenamefont
  {Fekete}, \citenamefont {Mazzera},\ and\ \citenamefont
  {de~Riedmatten}}]{Riedmatten2014}%
  \BibitemOpen
  \bibfield  {author} {\bibinfo {author} {\bibfnamefont {D.}~\bibnamefont
  {Riel\"ander}}, \bibinfo {author} {\bibfnamefont {K.}~\bibnamefont
  {Kutluer}}, \bibinfo {author} {\bibfnamefont {P.~M.}\ \bibnamefont
  {Ledingham}}, \bibinfo {author} {\bibfnamefont {M.}~\bibnamefont
  {G\"undo\ifmmode~\breve{g}\else \u{g}\fi{}an}}, \bibinfo {author}
  {\bibfnamefont {J.}~\bibnamefont {Fekete}}, \bibinfo {author} {\bibfnamefont
  {M.}~\bibnamefont {Mazzera}}, \ and\ \bibinfo {author} {\bibfnamefont
  {H.}~\bibnamefont {de~Riedmatten}},\ }\href {\doibase
  10.1103/PhysRevLett.112.040504} {\bibfield  {journal} {\bibinfo  {journal}
  {Phys. Rev. Lett.}\ }\textbf {\bibinfo {volume} {112}},\ \bibinfo {pages}
  {040504} (\bibinfo {year} {2014})}\BibitemShut {NoStop}%
\bibitem [{\citenamefont {Kolesov}\ \emph {et~al.}(2012)\citenamefont
  {Kolesov}, \citenamefont {Xia}, \citenamefont {Reuter}, \citenamefont
  {St\"{o}hr}, \citenamefont {Zappe}, \citenamefont {Meijer}, \citenamefont
  {Hemmer},\ and\ \citenamefont {Wrachtrup}}]{kolesov2012}%
  \BibitemOpen
  \bibfield  {author} {\bibinfo {author} {\bibfnamefont {R.}~\bibnamefont
  {Kolesov}}, \bibinfo {author} {\bibfnamefont {K.}~\bibnamefont {Xia}},
  \bibinfo {author} {\bibfnamefont {R.}~\bibnamefont {Reuter}}, \bibinfo
  {author} {\bibfnamefont {R.}~\bibnamefont {St\"{o}hr}}, \bibinfo {author}
  {\bibfnamefont {A.}~\bibnamefont {Zappe}}, \bibinfo {author} {\bibfnamefont
  {J.}~\bibnamefont {Meijer}}, \bibinfo {author} {\bibfnamefont
  {P.}~\bibnamefont {Hemmer}}, \ and\ \bibinfo {author} {\bibfnamefont
  {J.}~\bibnamefont {Wrachtrup}},\ }\href
  {http://dx.doi.org/10.1038/ncomms2034} {\bibfield  {journal} {\bibinfo
  {journal} {Nat Commun}\ }\textbf {\bibinfo {volume} {3}},\ \bibinfo {pages}
  {1029} (\bibinfo {year} {2012})}\BibitemShut {NoStop}%
\bibitem [{\citenamefont {Wisby}\ \emph {et~al.}(2014)\citenamefont {Wisby},
  \citenamefont {de~Graaf}, \citenamefont {Gwilliam}, \citenamefont {Adamyan},
  \citenamefont {Kubatkin}, \citenamefont {Meeson}, \citenamefont
  {Tzalenchuk},\ and\ \citenamefont {Lindstr\"om}}]{Wisby2014}%
  \BibitemOpen
  \bibfield  {author} {\bibinfo {author} {\bibfnamefont {I.}~\bibnamefont
  {Wisby}}, \bibinfo {author} {\bibfnamefont {S.~E.}\ \bibnamefont {de~Graaf}},
  \bibinfo {author} {\bibfnamefont {R.}~\bibnamefont {Gwilliam}}, \bibinfo
  {author} {\bibfnamefont {A.}~\bibnamefont {Adamyan}}, \bibinfo {author}
  {\bibfnamefont {S.~E.}\ \bibnamefont {Kubatkin}}, \bibinfo {author}
  {\bibfnamefont {P.~J.}\ \bibnamefont {Meeson}}, \bibinfo {author}
  {\bibfnamefont {A.~Y.}\ \bibnamefont {Tzalenchuk}}, \ and\ \bibinfo {author}
  {\bibfnamefont {T.}~\bibnamefont {Lindstr\"om}},\ }\href {\doibase
  http://dx.doi.org/10.1063/1.4894455} {\bibfield  {journal} {\bibinfo
  {journal} {Appl.~Phys.~Lett.}\ }\textbf {\bibinfo {volume} {105}},\ \bibinfo
  {eid} {102601} (\bibinfo {year} {2014})}\BibitemShut {NoStop}%
\bibitem [{\citenamefont {B\"{o}ttger}\ \emph {et~al.}(2006)\citenamefont
  {B\"{o}ttger}, \citenamefont {Thiel}, \citenamefont {Sun},\ and\
  \citenamefont {Cone}}]{Sun2006}%
  \BibitemOpen
  \bibfield  {author} {\bibinfo {author} {\bibfnamefont {T.}~\bibnamefont
  {B\"{o}ttger}}, \bibinfo {author} {\bibfnamefont {C.~W.}\ \bibnamefont
  {Thiel}}, \bibinfo {author} {\bibfnamefont {Y.}~\bibnamefont {Sun}}, \ and\
  \bibinfo {author} {\bibfnamefont {R.~L.}\ \bibnamefont {Cone}},\ }\href@noop
  {} {\bibfield  {journal} {\bibinfo  {journal} {Phys. Rev. B}\ }\textbf
  {\bibinfo {volume} {73}},\ \bibinfo {pages} {075101} (\bibinfo {year}
  {2006})}\BibitemShut {NoStop}%
\bibitem [{\citenamefont {Bertaina}\ \emph {et~al.}(2007)\citenamefont
  {Bertaina}, \citenamefont {Gambarelli}, \citenamefont {Tkachuk},
  \citenamefont {Kurkin}, \citenamefont {Malkin}, \citenamefont {Stepanov},\
  and\ \citenamefont {Barbara}}]{Bertaina2007}%
  \BibitemOpen
  \bibfield  {author} {\bibinfo {author} {\bibfnamefont {S.}~\bibnamefont
  {Bertaina}}, \bibinfo {author} {\bibfnamefont {S.}~\bibnamefont
  {Gambarelli}}, \bibinfo {author} {\bibfnamefont {A.}~\bibnamefont {Tkachuk}},
  \bibinfo {author} {\bibfnamefont {I.~N.}\ \bibnamefont {Kurkin}}, \bibinfo
  {author} {\bibfnamefont {B.}~\bibnamefont {Malkin}}, \bibinfo {author}
  {\bibfnamefont {A.}~\bibnamefont {Stepanov}}, \ and\ \bibinfo {author}
  {\bibfnamefont {B.}~\bibnamefont {Barbara}},\ }\href@noop {} {\bibfield
  {journal} {\bibinfo  {journal} {Nat.~Nanotechnol.}\ }\textbf {\bibinfo
  {volume} {2}},\ \bibinfo {pages} {39} (\bibinfo {year} {2007})}\BibitemShut
  {NoStop}%
\bibitem [{\citenamefont {Rakhmatullin}\ \emph {et~al.}(2009)\citenamefont
  {Rakhmatullin}, \citenamefont {Kurkin}, \citenamefont {Mamin}, \citenamefont
  {Orlinskii}, \citenamefont {Gafurov}, \citenamefont {Baibekov}, \citenamefont
  {Malkin}, \citenamefont {Gambarelli}, \citenamefont {Bertaina},\ and\
  \citenamefont {Barbara}}]{Bertaina2009PRB}%
  \BibitemOpen
  \bibfield  {author} {\bibinfo {author} {\bibfnamefont {R.~M.}\ \bibnamefont
  {Rakhmatullin}}, \bibinfo {author} {\bibfnamefont {I.~N.}\ \bibnamefont
  {Kurkin}}, \bibinfo {author} {\bibfnamefont {G.~V.}\ \bibnamefont {Mamin}},
  \bibinfo {author} {\bibfnamefont {S.~B.}\ \bibnamefont {Orlinskii}}, \bibinfo
  {author} {\bibfnamefont {M.~R.}\ \bibnamefont {Gafurov}}, \bibinfo {author}
  {\bibfnamefont {E.~I.}\ \bibnamefont {Baibekov}}, \bibinfo {author}
  {\bibfnamefont {B.~Z.}\ \bibnamefont {Malkin}}, \bibinfo {author}
  {\bibfnamefont {S.}~\bibnamefont {Gambarelli}}, \bibinfo {author}
  {\bibfnamefont {S.}~\bibnamefont {Bertaina}}, \ and\ \bibinfo {author}
  {\bibfnamefont {B.}~\bibnamefont {Barbara}},\ }\href@noop {} {\bibfield
  {journal} {\bibinfo  {journal} {Phys. Rev. B}\ }\textbf {\bibinfo {volume}
  {79}},\ \bibinfo {pages} {172408} (\bibinfo {year} {2009})}\BibitemShut
  {NoStop}%
\bibitem [{\citenamefont {Stannigel}\ \emph {et~al.}(2010)\citenamefont
  {Stannigel}, \citenamefont {Rabl}, \citenamefont {S{\o}rensen}, \citenamefont
  {Zoller},\ and\ \citenamefont {Lukin}}]{Rabl2010}%
  \BibitemOpen
  \bibfield  {author} {\bibinfo {author} {\bibfnamefont {K.}~\bibnamefont
  {Stannigel}}, \bibinfo {author} {\bibfnamefont {P.}~\bibnamefont {Rabl}},
  \bibinfo {author} {\bibfnamefont {A.~S.}\ \bibnamefont {S{\o}rensen}},
  \bibinfo {author} {\bibfnamefont {P.}~\bibnamefont {Zoller}}, \ and\ \bibinfo
  {author} {\bibfnamefont {M.~D.}\ \bibnamefont {Lukin}},\ }\href@noop {}
  {\bibfield  {journal} {\bibinfo  {journal} {Phys. Rev. Lett.}\ }\textbf
  {\bibinfo {volume} {105}},\ \bibinfo {pages} {220501} (\bibinfo {year}
  {2010})}\BibitemShut {NoStop}%
\bibitem [{\citenamefont {Tian}\ \emph {et~al.}(2004)\citenamefont {Tian},
  \citenamefont {Rabl}, \citenamefont {Blatt},\ and\ \citenamefont
  {Zoller}}]{Tian2004}%
  \BibitemOpen
  \bibfield  {author} {\bibinfo {author} {\bibfnamefont {L.}~\bibnamefont
  {Tian}}, \bibinfo {author} {\bibfnamefont {P.}~\bibnamefont {Rabl}}, \bibinfo
  {author} {\bibfnamefont {R.}~\bibnamefont {Blatt}}, \ and\ \bibinfo {author}
  {\bibfnamefont {P.}~\bibnamefont {Zoller}},\ }\href@noop {} {\bibfield
  {journal} {\bibinfo  {journal} {Phys. Rev. Lett.}\ }\textbf {\bibinfo
  {volume} {92}},\ \bibinfo {pages} {247902} (\bibinfo {year}
  {2004})}\BibitemShut {NoStop}%
\bibitem [{\citenamefont {Kukharchyk}\ \emph {et~al.}(2014)\citenamefont
  {Kukharchyk}, \citenamefont {Pal}, \citenamefont {R\"odiger}, \citenamefont
  {Ludwig}, \citenamefont {Probst}, \citenamefont {Ustinov}, \citenamefont
  {Bushev},\ and\ \citenamefont {Wieck}}]{Kukharchyk2014}%
  \BibitemOpen
  \bibfield  {author} {\bibinfo {author} {\bibfnamefont {N.}~\bibnamefont
  {Kukharchyk}}, \bibinfo {author} {\bibfnamefont {S.}~\bibnamefont {Pal}},
  \bibinfo {author} {\bibfnamefont {J.}~\bibnamefont {R\"odiger}}, \bibinfo
  {author} {\bibfnamefont {A.}~\bibnamefont {Ludwig}}, \bibinfo {author}
  {\bibfnamefont {S.}~\bibnamefont {Probst}}, \bibinfo {author} {\bibfnamefont
  {A.~V.}\ \bibnamefont {Ustinov}}, \bibinfo {author} {\bibfnamefont
  {P.}~\bibnamefont {Bushev}}, \ and\ \bibinfo {author} {\bibfnamefont {A.~D.}\
  \bibnamefont {Wieck}},\ }\href {\doibase 10.1002/pssr.201409304} {\bibfield
  {journal} {\bibinfo  {journal} {Phys. Status. Solidi (RRL)}\ ,\ \bibinfo
  {pages} {\textbf{DOI}~10.1002/pssr.201409304}} (\bibinfo {year}
  {2014})}\BibitemShut {NoStop}%
\bibitem [{\citenamefont {Melnikov}\ \emph {et~al.}(2002)\citenamefont
  {Melnikov}, \citenamefont {Gerya}, \citenamefont {Hillmann}, \citenamefont
  {Kamphausen}, \citenamefont {Oswald}, \citenamefont {Stauche}, \citenamefont
  {Wernhardt},\ and\ \citenamefont {Wieck}}]{melnikov2002}%
  \BibitemOpen
  \bibfield  {author} {\bibinfo {author} {\bibfnamefont {A.}~\bibnamefont
  {Melnikov}}, \bibinfo {author} {\bibfnamefont {T.}~\bibnamefont {Gerya}},
  \bibinfo {author} {\bibfnamefont {M.}~\bibnamefont {Hillmann}}, \bibinfo
  {author} {\bibfnamefont {I.}~\bibnamefont {Kamphausen}}, \bibinfo {author}
  {\bibfnamefont {W.}~\bibnamefont {Oswald}}, \bibinfo {author} {\bibfnamefont
  {P.}~\bibnamefont {Stauche}}, \bibinfo {author} {\bibfnamefont
  {R.}~\bibnamefont {Wernhardt}}, \ and\ \bibinfo {author} {\bibfnamefont
  {A.}~\bibnamefont {Wieck}},\ }\href@noop {} {\bibfield  {journal} {\bibinfo
  {journal} {Nuclear Instruments and Methods in Physics Research Section B:
  Beam Interactions with Materials and Atoms}\ }\textbf {\bibinfo {volume}
  {195}},\ \bibinfo {pages} {422} (\bibinfo {year} {2002})}\BibitemShut
  {NoStop}%
\bibitem [{\citenamefont {{Ziegler}}\ \emph {et~al.}(2010)\citenamefont
  {{Ziegler}}, \citenamefont {{Ziegler}},\ and\ \citenamefont
  {{Biersack}}}]{Ziegler2010}%
  \BibitemOpen
  \bibfield  {author} {\bibinfo {author} {\bibfnamefont {J.~F.}\ \bibnamefont
  {{Ziegler}}}, \bibinfo {author} {\bibfnamefont {M.~D.}\ \bibnamefont
  {{Ziegler}}}, \ and\ \bibinfo {author} {\bibfnamefont {J.~P.}\ \bibnamefont
  {{Biersack}}},\ }\href {\doibase 10.1016/j.nimb.2010.02.091} {\bibfield
  {journal} {\bibinfo  {journal} {Nucl. Instrum. and Methods in Phys. Res. B}\
  }\textbf {\bibinfo {volume} {268}},\ \bibinfo {pages} {1818} (\bibinfo {year}
  {2010})}\BibitemShut {NoStop}%
\bibitem [{\citenamefont {Schweiger}\ and\ \citenamefont
  {Eschke}(2001)}]{SchweigerESR}%
  \BibitemOpen
  \bibfield  {author} {\bibinfo {author} {\bibfnamefont {A.}~\bibnamefont
  {Schweiger}}\ and\ \bibinfo {author} {\bibfnamefont {G.}~\bibnamefont
  {Eschke}},\ }\href@noop {} {\emph {\bibinfo {title} {Principles of pulse
  elecrton paramagnetic resonance}}}\ (\bibinfo  {publisher} {Oxford University
  Press},\ \bibinfo {address} {Oxford, New York},\ \bibinfo {year}
  {2001})\BibitemShut {NoStop}%
\bibitem [{\citenamefont {Schuster}\ \emph {et~al.}(2010)\citenamefont
  {Schuster}, \citenamefont {Sears}, \citenamefont {Ginossar}, \citenamefont
  {DiCarlo}, \citenamefont {Frunzio}, \citenamefont {Morton}, \citenamefont
  {Wu}, \citenamefont {Briggs}, \citenamefont {Buckley}, \citenamefont
  {Awschalom},\ and\ \citenamefont {Schoelkopf}}]{Shuster2010}%
  \BibitemOpen
  \bibfield  {author} {\bibinfo {author} {\bibfnamefont {D.~I.}\ \bibnamefont
  {Schuster}}, \bibinfo {author} {\bibfnamefont {A.~P.}\ \bibnamefont {Sears}},
  \bibinfo {author} {\bibfnamefont {E.}~\bibnamefont {Ginossar}}, \bibinfo
  {author} {\bibfnamefont {L.}~\bibnamefont {DiCarlo}}, \bibinfo {author}
  {\bibfnamefont {L.}~\bibnamefont {Frunzio}}, \bibinfo {author} {\bibfnamefont
  {J.~J.~L.}\ \bibnamefont {Morton}}, \bibinfo {author} {\bibfnamefont
  {H.}~\bibnamefont {Wu}}, \bibinfo {author} {\bibfnamefont {G.~A.~D.}\
  \bibnamefont {Briggs}}, \bibinfo {author} {\bibfnamefont {B.~B.}\
  \bibnamefont {Buckley}}, \bibinfo {author} {\bibfnamefont {D.~D.}\
  \bibnamefont {Awschalom}}, \ and\ \bibinfo {author} {\bibfnamefont {R.~J.}\
  \bibnamefont {Schoelkopf}},\ }\href@noop {} {\bibfield  {journal} {\bibinfo
  {journal} {Phys. Rev. Lett.}\ }\textbf {\bibinfo {volume} {105}},\ \bibinfo
  {pages} {140501} (\bibinfo {year} {2010})}\BibitemShut {NoStop}%
\bibitem [{\citenamefont {Bushev}\ \emph {et~al.}(2011)\citenamefont {Bushev},
  \citenamefont {Feofanov}, \citenamefont {Rotzinger}, \citenamefont
  {Protopopov}, \citenamefont {Cole}, \citenamefont {Wilson}, \citenamefont
  {Fischer}, \citenamefont {Lukashenko},\ and\ \citenamefont
  {Ustinov}}]{Bushev2011}%
  \BibitemOpen
  \bibfield  {author} {\bibinfo {author} {\bibfnamefont {P.}~\bibnamefont
  {Bushev}}, \bibinfo {author} {\bibfnamefont {A.~K.}\ \bibnamefont
  {Feofanov}}, \bibinfo {author} {\bibfnamefont {H.}~\bibnamefont {Rotzinger}},
  \bibinfo {author} {\bibfnamefont {I.}~\bibnamefont {Protopopov}}, \bibinfo
  {author} {\bibfnamefont {J.~H.}\ \bibnamefont {Cole}}, \bibinfo {author}
  {\bibfnamefont {C.~M.}\ \bibnamefont {Wilson}}, \bibinfo {author}
  {\bibfnamefont {G.}~\bibnamefont {Fischer}}, \bibinfo {author} {\bibfnamefont
  {A.}~\bibnamefont {Lukashenko}}, \ and\ \bibinfo {author} {\bibfnamefont
  {A.~V.}\ \bibnamefont {Ustinov}},\ }\href@noop {} {\bibfield  {journal}
  {\bibinfo  {journal} {Phys. Rev. B}\ }\textbf {\bibinfo {volume} {84}},\
  \bibinfo {pages} {060501(R)} (\bibinfo {year} {2011})}\BibitemShut {NoStop}%
\bibitem [{\citenamefont {Kubo}\ \emph {et~al.}(2012)\citenamefont {Kubo},
  \citenamefont {Diniz}, \citenamefont {Grezes}, \citenamefont {Umeda},
  \citenamefont {Isoya}, \citenamefont {Sumiya}, \citenamefont {Yamamoto},
  \citenamefont {Abe}, \citenamefont {Onoda}, \citenamefont {Ohshima},
  \citenamefont {Jacques}, \citenamefont {Dr\'eau}, \citenamefont {Roch},
  \citenamefont {Auffeves}, \citenamefont {Vion}, \citenamefont {Esteve},\ and\
  \citenamefont {Bertet}}]{Bertet2012}%
  \BibitemOpen
  \bibfield  {author} {\bibinfo {author} {\bibfnamefont {Y.}~\bibnamefont
  {Kubo}}, \bibinfo {author} {\bibfnamefont {I.}~\bibnamefont {Diniz}},
  \bibinfo {author} {\bibfnamefont {C.}~\bibnamefont {Grezes}}, \bibinfo
  {author} {\bibfnamefont {T.}~\bibnamefont {Umeda}}, \bibinfo {author}
  {\bibfnamefont {J.}~\bibnamefont {Isoya}}, \bibinfo {author} {\bibfnamefont
  {H.}~\bibnamefont {Sumiya}}, \bibinfo {author} {\bibfnamefont
  {T.}~\bibnamefont {Yamamoto}}, \bibinfo {author} {\bibfnamefont
  {H.}~\bibnamefont {Abe}}, \bibinfo {author} {\bibfnamefont {S.}~\bibnamefont
  {Onoda}}, \bibinfo {author} {\bibfnamefont {T.}~\bibnamefont {Ohshima}},
  \bibinfo {author} {\bibfnamefont {V.}~\bibnamefont {Jacques}}, \bibinfo
  {author} {\bibfnamefont {A.}~\bibnamefont {Dr\'eau}}, \bibinfo {author}
  {\bibfnamefont {J.-F.}\ \bibnamefont {Roch}}, \bibinfo {author}
  {\bibfnamefont {A.}~\bibnamefont {Auffeves}}, \bibinfo {author}
  {\bibfnamefont {D.}~\bibnamefont {Vion}}, \bibinfo {author} {\bibfnamefont
  {D.}~\bibnamefont {Esteve}}, \ and\ \bibinfo {author} {\bibfnamefont
  {P.}~\bibnamefont {Bertet}},\ }\href {\doibase 10.1103/PhysRevB.86.064514}
  {\bibfield  {journal} {\bibinfo  {journal} {Phys. Rev. B}\ }\textbf {\bibinfo
  {volume} {86}},\ \bibinfo {pages} {064514} (\bibinfo {year}
  {2012})}\BibitemShut {NoStop}%
\bibitem [{\citenamefont {Sigillito}\ \emph {et~al.}(2014)\citenamefont
  {Sigillito}, \citenamefont {Malissa}, \citenamefont {Tyryshkin},
  \citenamefont {Riemann}, \citenamefont {Abrosimov}, \citenamefont {Becker},
  \citenamefont {Pohl}, \citenamefont {Thewalt}, \citenamefont {Itoh},
  \citenamefont {Morton}, \citenamefont {Houck}, \citenamefont {Schuster},\
  and\ \citenamefont {Lyon}}]{Morton2013}%
  \BibitemOpen
  \bibfield  {author} {\bibinfo {author} {\bibfnamefont {A.~J.}\ \bibnamefont
  {Sigillito}}, \bibinfo {author} {\bibfnamefont {H.}~\bibnamefont {Malissa}},
  \bibinfo {author} {\bibfnamefont {A.~M.}\ \bibnamefont {Tyryshkin}}, \bibinfo
  {author} {\bibfnamefont {H.}~\bibnamefont {Riemann}}, \bibinfo {author}
  {\bibfnamefont {N.~V.}\ \bibnamefont {Abrosimov}}, \bibinfo {author}
  {\bibfnamefont {P.}~\bibnamefont {Becker}}, \bibinfo {author} {\bibfnamefont
  {H.-J.}\ \bibnamefont {Pohl}}, \bibinfo {author} {\bibfnamefont {M.~L.~W.}\
  \bibnamefont {Thewalt}}, \bibinfo {author} {\bibfnamefont {K.~M.}\
  \bibnamefont {Itoh}}, \bibinfo {author} {\bibfnamefont {J.~J.~L.}\
  \bibnamefont {Morton}}, \bibinfo {author} {\bibfnamefont {A.~A.}\
  \bibnamefont {Houck}}, \bibinfo {author} {\bibfnamefont {D.~I.}\ \bibnamefont
  {Schuster}}, \ and\ \bibinfo {author} {\bibfnamefont {S.~A.}\ \bibnamefont
  {Lyon}},\ }\href {\doibase http://dx.doi.org/10.1063/1.4881613} {\bibfield
  {journal} {\bibinfo  {journal} {Appl.~Phys.~Lett.}\ }\textbf {\bibinfo
  {volume} {104}},\ \bibinfo {eid} {222407} (\bibinfo {year}
  {2014})}\BibitemShut {NoStop}%
\bibitem [{\citenamefont {Brown}\ \emph {et~al.}(2011)\citenamefont {Brown},
  \citenamefont {Tyryshkin}, \citenamefont {Porfyrakis}, \citenamefont
  {Gauger}, \citenamefont {Lovett}, \citenamefont {Ardavan}, \citenamefont
  {Lyon}, \citenamefont {Briggs},\ and\ \citenamefont {Morton}}]{Morton2011}%
  \BibitemOpen
  \bibfield  {author} {\bibinfo {author} {\bibfnamefont {R.~M.}\ \bibnamefont
  {Brown}}, \bibinfo {author} {\bibfnamefont {A.~M.}\ \bibnamefont
  {Tyryshkin}}, \bibinfo {author} {\bibfnamefont {K.}~\bibnamefont
  {Porfyrakis}}, \bibinfo {author} {\bibfnamefont {E.~M.}\ \bibnamefont
  {Gauger}}, \bibinfo {author} {\bibfnamefont {B.~W.}\ \bibnamefont {Lovett}},
  \bibinfo {author} {\bibfnamefont {A.}~\bibnamefont {Ardavan}}, \bibinfo
  {author} {\bibfnamefont {S.~A.}\ \bibnamefont {Lyon}}, \bibinfo {author}
  {\bibfnamefont {G.~A.~D.}\ \bibnamefont {Briggs}}, \ and\ \bibinfo {author}
  {\bibfnamefont {J.~J.~L.}\ \bibnamefont {Morton}},\ }\href {\doibase
  10.1103/PhysRevLett.106.110504} {\bibfield  {journal} {\bibinfo  {journal}
  {Phys. Rev. Lett.}\ }\textbf {\bibinfo {volume} {106}},\ \bibinfo {pages}
  {110504} (\bibinfo {year} {2011})}\BibitemShut {NoStop}%
\bibitem [{\citenamefont {Bhattacharyya}\ and\ \citenamefont
  {Chakrabarti}(2008)}]{Bhattacharyya2008}%
  \BibitemOpen
  \bibfield  {author} {\bibinfo {author} {\bibfnamefont {P.}~\bibnamefont
  {Bhattacharyya}}\ and\ \bibinfo {author} {\bibfnamefont {B.~K.}\ \bibnamefont
  {Chakrabarti}},\ }\href {http://stacks.iop.org/0143-0807/29/i=3/a=023}
  {\bibfield  {journal} {\bibinfo  {journal} {Eur.~J.~Phys.}\ }\textbf
  {\bibinfo {volume} {29}},\ \bibinfo {pages} {639} (\bibinfo {year}
  {2008})}\BibitemShut {NoStop}%
\bibitem [{\citenamefont {Abragam}\ and\ \citenamefont
  {Bleaney}(2012)}]{AbragamESR}%
  \BibitemOpen
  \bibfield  {author} {\bibinfo {author} {\bibfnamefont {A.}~\bibnamefont
  {Abragam}}\ and\ \bibinfo {author} {\bibfnamefont {B.}~\bibnamefont
  {Bleaney}},\ }\href@noop {} {\emph {\bibinfo {title} {Electron paramagnetic
  resonance of transition ions}}}\ (\bibinfo  {publisher} {Oxford University
  Press},\ \bibinfo {address} {Oxford},\ \bibinfo {year} {2012})\BibitemShut
  {NoStop}%
\bibitem [{\citenamefont {Wuensch}\ \emph {et~al.}(2011)\citenamefont
  {Wuensch}, \citenamefont {Hammer}, \citenamefont {Kappler}, \citenamefont
  {Geupert},\ and\ \citenamefont {Siegel}}]{Wuensch2011}%
  \BibitemOpen
  \bibfield  {author} {\bibinfo {author} {\bibfnamefont {S.}~\bibnamefont
  {Wuensch}}, \bibinfo {author} {\bibfnamefont {G.}~\bibnamefont {Hammer}},
  \bibinfo {author} {\bibfnamefont {T.}~\bibnamefont {Kappler}}, \bibinfo
  {author} {\bibfnamefont {F.}~\bibnamefont {Geupert}}, \ and\ \bibinfo
  {author} {\bibfnamefont {M.}~\bibnamefont {Siegel}},\ }\href@noop {}
  {\bibfield  {journal} {\bibinfo  {journal} {IEEE Transactions on Appl.
  Superconductivity}\ }\textbf {\bibinfo {volume} {21}},\ \bibinfo {pages}
  {752} (\bibinfo {year} {2011})}\BibitemShut {NoStop}%
\bibitem [{\citenamefont {Schuck}\ \emph {et~al.}(2013)\citenamefont {Schuck},
  \citenamefont {Pernice},\ and\ \citenamefont {Tang}}]{Pernice2013}%
  \BibitemOpen
  \bibfield  {author} {\bibinfo {author} {\bibfnamefont {C.}~\bibnamefont
  {Schuck}}, \bibinfo {author} {\bibfnamefont {W.~H.~P.}\ \bibnamefont
  {Pernice}}, \ and\ \bibinfo {author} {\bibfnamefont {H.~X.}\ \bibnamefont
  {Tang}},\ }\href {\doibase http://dx.doi.org/10.1063/1.4788931} {\bibfield
  {journal} {\bibinfo  {journal} {Appl.~Phys.~Lett.}\ }\textbf {\bibinfo
  {volume} {102}},\ \bibinfo {eid} {051101} (\bibinfo {year}
  {2013})}\BibitemShut {NoStop}%
\bibitem [{\citenamefont {O'Brien}\ \emph {et~al.}(2014)\citenamefont
  {O'Brien}, \citenamefont {Lauk}, \citenamefont {Blum}, \citenamefont
  {Morigi},\ and\ \citenamefont {Fleischhauer}}]{Morigi2014}%
  \BibitemOpen
  \bibfield  {author} {\bibinfo {author} {\bibfnamefont {C.}~\bibnamefont
  {O'Brien}}, \bibinfo {author} {\bibfnamefont {N.}~\bibnamefont {Lauk}},
  \bibinfo {author} {\bibfnamefont {S.}~\bibnamefont {Blum}}, \bibinfo {author}
  {\bibfnamefont {G.}~\bibnamefont {Morigi}}, \ and\ \bibinfo {author}
  {\bibfnamefont {M.}~\bibnamefont {Fleischhauer}},\ }\href@noop {} {\bibfield
  {journal} {\bibinfo  {journal} {Phys. Rev. Lett.}\ }\textbf {\bibinfo
  {volume} {113}},\ \bibinfo {pages} {063603} (\bibinfo {year}
  {2014})}\BibitemShut {NoStop}%
\end{thebibliography}%

\end{document}